\documentclass[a4paper,fleqn,usenatbib,useAMS]{mnras}

\usepackage[T1]{fontenc}
\usepackage{ae,aecompl}

\usepackage{graphicx}	
\usepackage{amsmath}	
\usepackage{amssymb}	
\usepackage{hyperref}
\usepackage{bm}
\renewcommand{\theenumi}{\arabic{enumi}.}
\title[Parsec-scale jet properties: PG 1302$-$102]{Parsec-scale jet properties of the quasar PG 1302$-$102}
\author[Mohan et al.]
{P. Mohan$^{1}$\thanks{E-mail: pmohan@shao.ac.cn}, 
T. An$^{1,2}$\thanks{E-mail: antao@shao.ac.cn},
S. Frey$^{3}$,
A. Mangalam$^{4}$,
K.~{\'E}. Gab{\'a}nyi$^{3,5}$ and
E. Kun$^{6,7}$.\\ \\
$^{1}$Shanghai Astronomical Observatory, 80 Nandan Road, Shanghai 200030, China\\
$^{2}$Key Laboratory of Radio Astronomy, Chinese Academy of Sciences, Nanjing 210008, China\\
$^{3}$F{\"O}MI, Satellite Geodetic Observatory, P.O. Box 585, H-1592 Budapest, Hungary\\
$^{4}$Indian Institute of Astrophysics, Sarjapur Road, Koramangala, Bangalore 560034, India\\
$^{5}$Konkoly Observatory, MTA Research Centre for Astronomy and Earth Sciences, P.O. Box 67, 1525 Budapest, Hungary\\
$^{6}$Department of Experimental Physics, University of Szeged, D{\'o}m t{\'e}r 9, H-6720 Szeged, Hungary\\
$^{7}$Department of Theoretical Physics, University of Szeged, Tisza Lajos krt 84-86, H-6720 Szeged, Hungary
}

\date{Accepted 2016 August 23. Received 2016 August 23; in original form 2016 July 14}

\pubyear{2016}

\begin{document}
\label{firstpage}
\pagerange{\pageref{firstpage}--\pageref{lastpage}}
\maketitle

\begin{abstract}
The quasar PG 1302$-$102 is believed to harbour a supermassive binary black hole (SMBBH) system. Using the available 15 GHz and $2-8$ GHz, multi-epoch Very Long Baseline Array data, we constrain the pc-scale jet properties based on the inferred mean proper motion, including a bulk Lorentz factor $\geqslant 5.1 \pm 0.8$, jet inclination angle $\leqslant 11\fdg4 \pm 1\fdg7$, projected position angle $= 31\fdg8$, intrinsic half opening angle $\leqslant 0\fdg9 \pm 0\fdg1$ and a mean $2-8$ GHz spectral index of 0.31. A general relativistic helical jet model is presented and applied to predict quasi-periodic oscillations of $\sim$ 10 days, power law power spectrum shape and a contribution of up to $\sim$ 53 percent to the observed variable core flux density. The model is used to make a case for high resolution, moderately sampled, long duration radio interferometric observations to reveal signatures due to helical knots and distinguish them from those due to SMBBH orbital activity including a phase difference $\sim \pi$ and an amplitude ratio (helical light curve amplitude/SMBBH light curve amplitude) of $0.2-3.3$. The prescription can be used to identify helical kinematic signatures from quasars, providing possible candidates for further studies with polarization measurements. It can also be used to infer promising SMBBH candidates for the study of gravitational waves if there are systematic deviations from helical signatures.
\end{abstract}


\section{Introduction}

Recent intensive studies of the quasar PG 1302$-$102 have brought into focus the possible ubiquity of supermassive binary black hole (SMBBH) systems powering active galactic nuclei (AGN) \cite[e.g.][]{1980Natur.287..307B}, which is expected in the context of hierarchical models of cosmological evolution \cite[e.g.][]{2003ApJ...582..559V,2015ApJ...799..178K} as well as in galaxy mergers \cite[e.g.][]{2016arXiv160606568K} including simulations \cite[e.g.][]{2001ApJ...558..535M,2015MNRAS.449..494R}. The detection of a SMBBH system has been approached through multiple observational and theoretical schemes, including the search for kpc-separated dual AGN \cite[e.g.][]{2003ApJ...582L..15K,2008MNRAS.386..105B,2012ApJ...753...42C,2014MNRAS.437...32W}, which could be in a merging state; follow-up studies of identified double-peaked narrow emission lines from AGN, by an optical spectral analysis \cite[e.g.][]{2009ApJ...705L..20X}, and very long baseline interferometry (VLBI) observations of the radio jet morphology \cite[e.g.][]{2012MNRAS.425.1185F,2013MNRAS.433.1161A,2014MNRAS.445.1370K,2014Natur.511...57D,2016A&A...588A.102B,2016ApJ...826..106G}; quasi-periodic variations in optical flux \cite[e.g.][]{2008Natur.452..851V,2015Natur.518...74G,2015MNRAS.453.1562G,2015ApJ...803L..16L,2016arXiv160401020C}, ultraviolet \citep{2015Natur.525..351D}, and in the optical continuum and emission lines \citep{2016ApJ...822....4L}; unique signatures in the optical to ultraviolet spectrum \citep{2015ApJ...809..117Y}, expected to emerge from sub-parsec binary activity; and, the drawing of observable inferences from simulations of the SMBBH involving the accretion disk, jet, and their interactions \cite[e.g.][]{2016MNRAS.455.1989G}.

An important space with a strong potential for novel discoveries has been opened recently with the detection of gravitational waves from the coalescence of two stellar-mass black holes ($M_\bullet \sim 30 M_\odot$) by the Laser Interferometer Gravitational-wave Observatory \cite[LIGO; ][]{2016PhRvL.116f1102A}. The SMBBH ($M_\bullet = 10^8 - 10^{10} M_\odot$) coalescence in galaxies at redshift $z \leqslant 1.5$ is expected to produce a gravitational wave background with a strain amplitude of $\sim 10^{-15} - 10^{-16}$ at the nano-Hz frequencies \citep{2015MNRAS.447.2772R,2016ApJ...821...13A} which may not be detected by LIGO owing to its operational sensitivity tuned for detection at high frequencies (e.g. $10 - 10^3$ Hz). However, this background can be detected using pulsar timing arrays \cite[PTAs; e.g.][]{2008MNRAS.390..192S,2010CQGra..27h4013H}, which employ a highly accurate estimate of the pulse arrival times from a growing number of stable pulsars in order to detect any changes in strain amplitudes resulting from the gravitational wave signals at the nHz frequencies. In addition, the proposed European Laser Interferometer Space Antenna \cite[e-LISA; ][]{2012CQGra..29l4016A} is sensitive to detect gravitational waves in the $(10^{-4}-1)$ Hz frequency range which is expected to be dominated by the inspiral, merger and ringdown events from lower mass ($M_\bullet = 10^6 - 10^{8} M_\odot$) SMBBH systems \cite[e.g.][]{2016PhRvD..93b4003K}.

The quasar PG 1302$-$102, at a redshift $z = 0.2784$ \citep{1996ApJS..104...37M} 
was identified as a strongly optically variable source with a possible periodicity of 1884 $\pm$ 88 d (observer frame) in a systematic time series analysis of the light curves spanning $\sim$ 9 y from 247,000 spectroscopically identified quasars in the Catalina Real-time Transient Survey \cite[CRTS; ][]{2015Natur.518...74G}. The study discusses the origin of the optical periodicity in the context of jet precession, a warped accretion disk and periodic accretion and argues for its association with the orbital motion of a SMBBH with separation $d \leq 0.01$ pc. 
Motivated by hydrodynamic simulations of circum-binary disks in an unequal-mass (black hole mass ratio $q = M_1/M_2 \leq 0.1$) SMBBH system, the study of \cite{2015Natur.525..351D} indicates that the circumsecondary disk dominates the total flux and the variability, and relativistic beaming provides an explanation for the optical periodicity. The observed variability is attributed to the orbital motion of an over-dense region in the circum-binary disk for a near equal mass ($0.3 \leq q \leq 0.8$) SMBBH system \citep{2015MNRAS.452.2540D}, where the orbital period of this region is $(3-8)$ times the binary orbital period, leading to a reduced separation, $d \sim 0.003$ pc. A statistical analysis of the compiled optical light curve is carried out in \cite{2015MNRAS.454L..21C}, finding no evidence for statistically significant harmonics of the observed optical periodicity, which are expected from hydrodynamic simulations. The study of \cite{2015ApJ...814L..12J} presents the 3.4 and 4.6 $\mu$m infrared archival light curves and detects a time lag from the optical emission, consistent with it being re-processed by the AGN torus and implying that the optical periodicity is of disk origin. The pc-scale and kpc-scale radio properties of PG 1302$-$102 jet are studied by \cite{2015MNRAS.454.1290K} to address the jet variability, its morphological properties, the emission of gravitational waves during the inspiral and associated timescales, also obtaining a constraint of $q \geqslant 0.08$. Assuming the beamed jet model scenario proposed by \cite{2015Natur.518...74G}, they obtain similar kinematic parameters which are discussed and compared in the following section.
At the kpc-scale, a two-sided, asymmetric jet with wider intrinsic half opening angle and inclination angle with possible expanding helical structure are inferred.  

Our paper is organized as follows: in Section \ref{pcjet}, we estimate kinematic parameters and spectral indices for the pc-scale jet of PG 1302$-$102; in Section \ref{kinjet}, we present the helical jet model and describe its applicability; in Section \ref{Modelapp}, we simulate the expected light curves and component trajectories for PG 1302$-$102. The model is applied to infer helical signatures in the variable core flux density. The expected variability from helical kinematics is then distinguished from that arising in a SMBBH scenario; in Section \ref{conclusion}, the results are presented and the applicability of the analysis is discussed in the context of being a preliminary search for helical signatures (possibly due to magnetic fields) and in aiding in the discovery of SMBBH systems.

\section{Parsec-scale jet properties}
\label{pcjet}

\begin{figure}
\centerline{\includegraphics[scale=0.26]{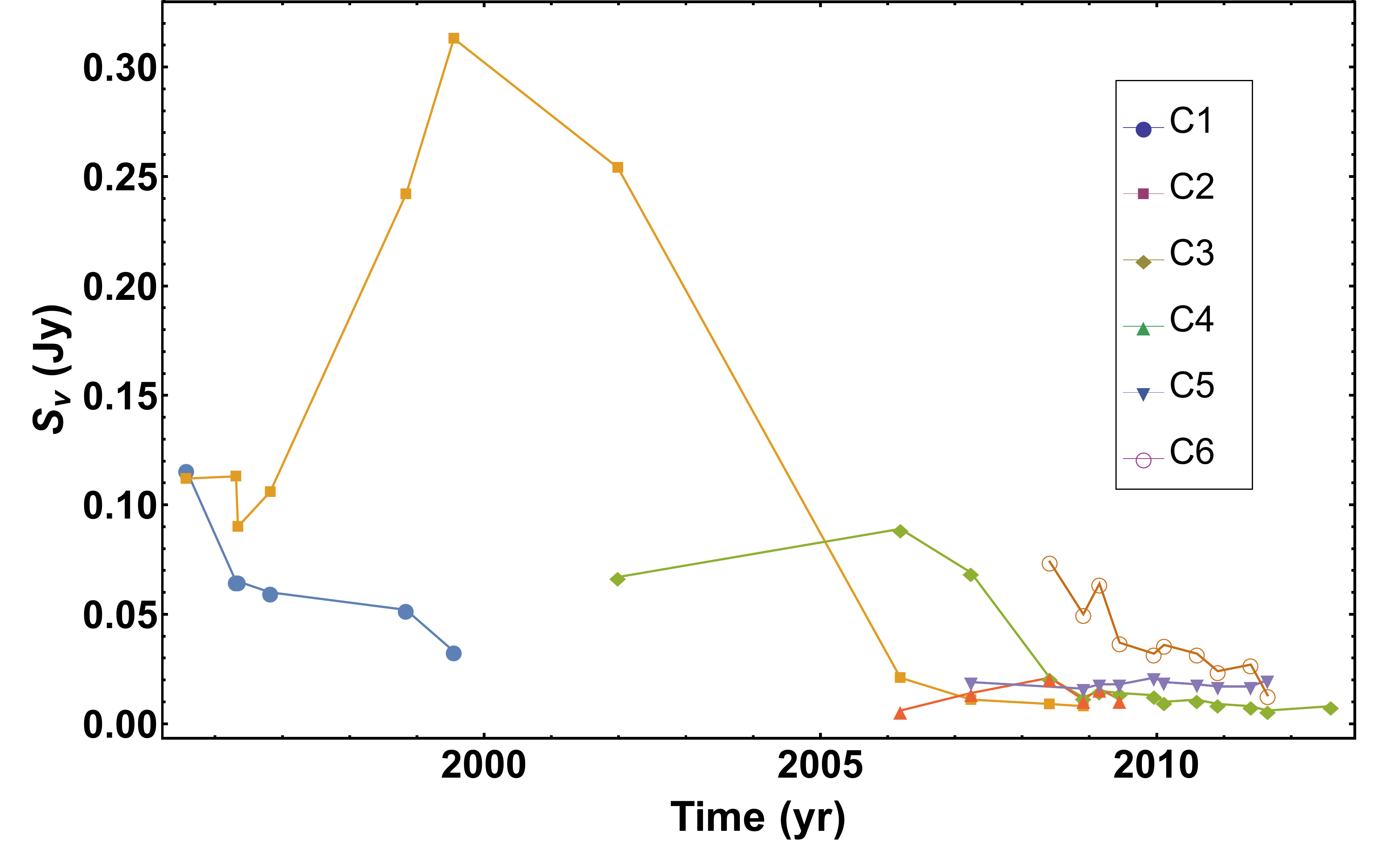}}
\caption{The 15 GHz flux density ($S_\nu$) variation with observation time for individual VLBI components C1--C6.}
\label{Snuobs}
\end{figure}

The kinematic and flux density variability data used in the following analysis are that reduced by \cite{2015MNRAS.454.1290K} using calibrated VLBI visibility data of PG 1302$-$102 from the MOJAVE 15 GHz survey \citep{2009AJ....138.1874L} using the Very Large Baseline Array (VLBA), spanning 20 epochs between $1995-2012$. The VLBI images obtained during the snapshot observations \citep{2013AJ....146..120L} can be used to study the pc-scale evolution over the observation epochs\footnote{All MOJAVE images and calibrated visibility data of PG 1302$-$102 are available from \url{http://www.physics.purdue.edu/astro/MOJAVE/sourcepages/1302-102.shtml}}. Six resolved bright components C1--C6 are identified in the pc-scale jet from model fitting. For details on the data reduction procedure, including the Gaussian model fitting, core and component identification and extraction of their positions and flux density, we refer the reader to the discussion in Section 2 of \cite{2015MNRAS.454.1290K}. Our re-analysis of these data here serves the purpose of estimating the jet kinematic parameters in an independent manner (in the earlier study, the Doppler factor and subsequent kinematic parameters were estimated from the measured brightness temperature), and to further the discussion on the pc-scale jet properties. The flux density ($S_\nu$) variation of the components is presented in Fig. \ref{Snuobs}. 

For right ascension R.A. (in mas) and declination DEC. (in mas) relative to the core position, we fit a linear function to the core separation $\omega = ({\rm R.A.}^2+{\rm DEC.}^2)^{1/2}$ (mas) as a function of observation time $\tilde{\tau}$ (in yr) for each component to obtain their proper motion $\mu = d\omega/d\tilde{\tau}$ (mas yr$^{-1}$), in the range $(0.17 - 0.41)$ mas yr$^{-1}$ with a mean of $0.27 \pm 0.04$ mas yr$^{-1}$, and presented in Table \ref{tab1}. 

Assuming a flat cold dark matter dominated cosmology with matter energy density $\Omega_{m} = 0.308$ and Hubble constant $H_0 = 67.8~$km s$^{-1}$ Mpc$^{-1}$ \cite[][]{2015arXiv150201589P}, the scale at $z = 0.2784$ is 4.357 kpc arcsec$^{-1}$ and the corresponding luminosity distance $D_L = 1468.7$ Mpc \citep{2006PASP..118.1711W}. We derive the jet properties assuming a conically shaped jet, supporting a bulk relativistic flow with constant opening angle and speed. For the mean proper motion $\mu = 0.27 \pm 0.04$ mas yr$^{-1}$, the transverse speed (apparent superluminal speed in units of the speed of light $c$) is $\beta_\perp = 5.0\pm 0.8$, the corresponding jet bulk speed $\beta \geqslant \beta_\perp/(1+\beta^2_\perp)^{1/2} = 0.980 \pm 0.006$, and bulk Lorentz factor $\Gamma = (1-\beta^2)^{-1/2} \geqslant 5.1 \pm 0.8$ (equal to the Doppler factor $D$). The associated inclination angle to the observer line of sight $i \leqslant \cos^{-1} \beta = 11\fdg4 \pm 1\fdg7$. 

\begin{figure}
\centerline{\includegraphics[scale=0.27]{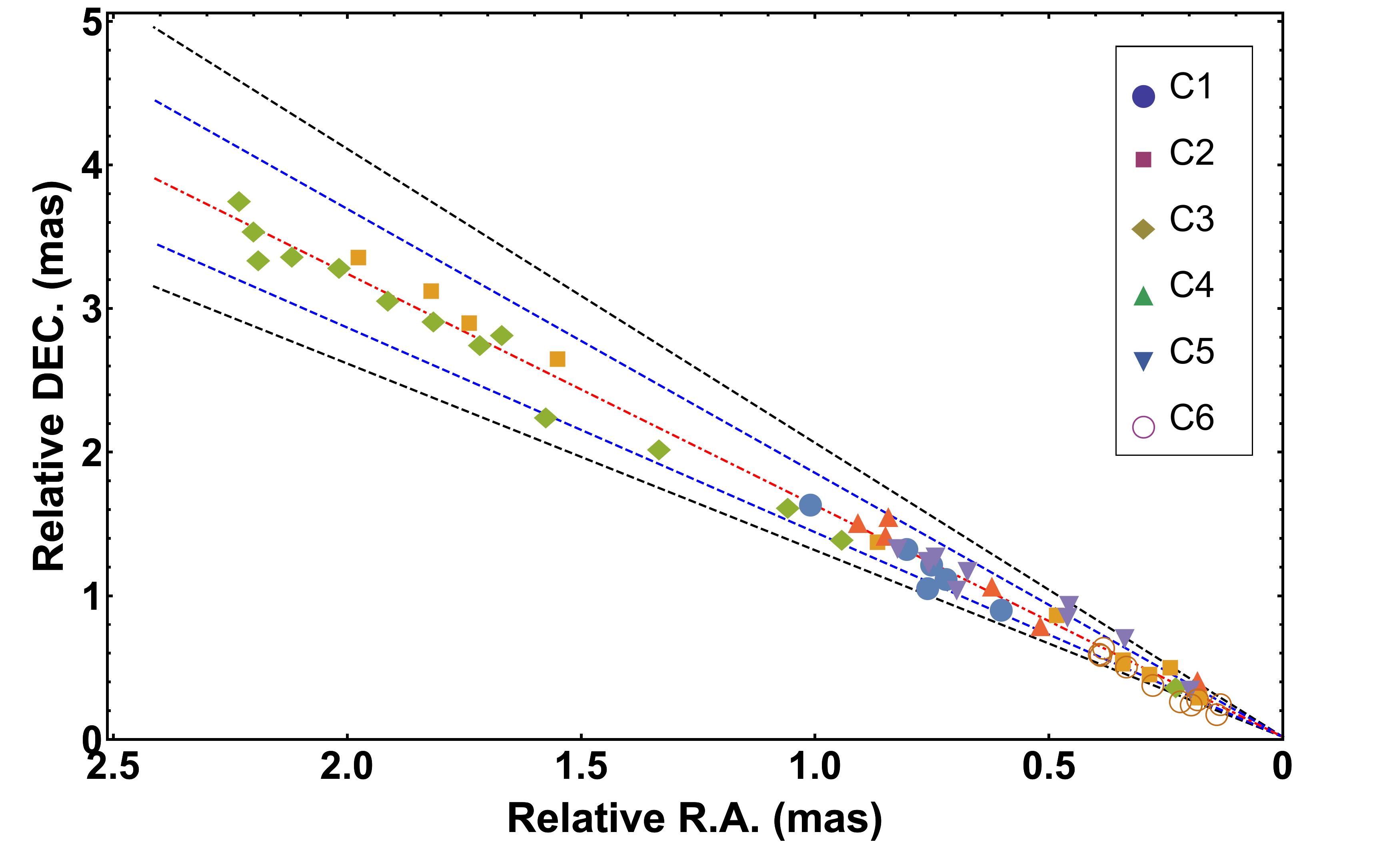}}
\caption{The jet half opening angle is estimated from the component $({\rm R.A.,DEC.})$ distribution. The mean jet axis is the red dot-dashed line. The projected half opening angle from the outer boundaries (black, dashed lines) is $\psi = 5\fdg8$ and from the inner boundaries (blue dashed lines) is $3\fdg3$ with a mean of $\psi = 4\fdg5 \pm 1\fdg3$.} 
\label{hoa}
\end{figure}

We then determine the angle enclosing all jet components and hence, the projected and intrinsic jet half opening angle by fitting the relative DEC. versus R.A. of all components, presented in Fig. \ref{hoa}. A linear fit to the relative DEC. versus R.A. data gives the mean jet axis direction with a normalization of $0.017$ and a slope of $1.61$ from which, the mean projected position angle $\lambda = 31\fdg8 \pm 0\fdg2$, here defined to be clockwise from the $Y$-axis (DEC.). With the fit residuals, we determine the position of minimum and maximum deviations from the jet axis, used to prescribe the jet boundaries. The projected half opening angle for the inner and outer boundaries are $3\fdg3$ and $5\fdg8$ respectively with a mean $\psi = 4\fdg5 \pm 1\fdg3$. The intrinsic half opening angle is $\theta_0 = \psi \sin i \leqslant 0\fdg9 \pm 0\fdg1$. The inferred kinematic parameters are summarized in Table \ref{tab1}. The study of \cite{2004ApJ...609..539K} identifies a composite component in three snapshot observations of PG 1302-102 between 1995--1999, giving $\lambda = 30^{\circ}$, $\mu = 0.31 \pm 0.05$ mas yr$^{-1}$ and $\beta_\perp = 5.6 \pm 0.9$. The study of \cite{2015MNRAS.454.1290K} identifies six components in the multi-epoch observations and, from an assumed maximum equipartition brightness temperature (intrinsic temperature) of $5 \times 10^{10}$ K, determines a median $D = 18.5 \pm 3.6$, $\Gamma = 10.8^{+1.7}_{-1.9}$, $\lambda = 31\fdg6 \pm 0\fdg6$, $\theta_0 = 0\fdg20 \pm 0\fdg05$ and $i = 2\fdg2 \pm 0\fdg5$. The study of \cite{2016AJ....152...12L} identifies five components in the multi-epoch observations, giving $\lambda = (29\fdg5 - 32\fdg9)$, $\mu = (0.18 \pm 0.02 - 0.49 \pm 0.02)$ mas yr$^{-1}$ and $\beta_\perp = (3.1 \pm 0.4 - 8.5 \pm 0.3)$. Our parameter estimates are within the range reported by these earlier studies. The estimated $\Gamma$, $\theta_0$ and $i$ differ but as they are evaluated as limits, using only the kinematical information as opposed to the calculation of $D$ and subsequent parameters from the apparent brightness temperatures in \cite{2015MNRAS.454.1290K}.

In Fig. \ref{hoa}, all components are within the outer boundary till $({\rm R.A.,DEC.}) \sim (0.5,1)$ mas and are clustered closer to the mean beyond this region, and contained within the inner boundary. The components C2, C3 and C4 which are mainly contained within the inner boundaries in the region $({\rm R.A.},{\rm DEC.}) \geqslant (0.5,1)$ mas are associated with higher proper motions. 
If we identify C4 with the component no. 4 of \cite{2016AJ....152...12L} (based on its high proper motion; correspondence with their Tables 4 and 5 including a mean separation of 3.39 mas), it is inferred to decelerate with $\sim$ 49 $\mu$as yr$^{-2}$ in the direction parallel to the apparent motion indicating possible slowdown due to ballistic motion or interaction with the interstellar medium, resulting in the region beyond $({\rm R.A.},{\rm DEC.}) \sim (1.1,1.8)$ mas being populated only by the components C2 and C3. 

The source was also observed with the VLBA simultaneously in the S band ($\sim$ 2 GHz) and X band ($\sim$ 8 GHz) during five epochs spanning between $1997 - 2005$.
We fitted single elliptical Gaussian brightness distribution model components to the visibility data\footnote{The calibrated data sets were downloaded from the Astrogeo Database maintained by L. Petrov, \url{http://www.astrogeo.org}} with DIFMAP \cite[][]{1997ASPC..125...77S}. The model fitting results are presented in Table \ref{tab2} which includes the observation epoch, the central observing frequency, flux density, fitted component major axis $\theta_{\rm maj}$, axis ratio, component position angle (P.A.), and the spectral index, calculated assuming that the flux density $S_\nu$ and observing frequency $\nu$ are related by $S_\nu \propto \nu^{-\alpha}$. The inferred spectral index $\alpha$ ranges between $0.18 - 0.48$, indicating a flat spectrum source with a mean $\bar{\alpha} = 0.31$. There is a hint of steepening of $\alpha$ with observation epoch and total flux density indicating that the optically-thin jet  based contribution could be significant during later epochs. Though, with the available sparsely sampled data and the lack of additional simultaneous multiband data, it is difficult to track the evolution and infer whether it is due to any underlying physical process.

\begin{table}
\scalebox{.9}{
\begin{tabular}{lll}
\hline
Kinematic quantity & Symbol & Estimate \\ \hline \hline
Component proper motion (mas yr$^{-1}$) & C1 & $0.17\pm0.03$ \\
                    & C2 & $0.26\pm0.01$ \\
                    & C3 & $0.39\pm0.01$ \\
                    & C4 & $0.41\pm0.03$ \\
                    & C5 & $0.24\pm0.02$ \\
                    & C6 & $0.17\pm0.02$ \\
Apparent speed (units of $c$) & $\beta_{\perp}$ & $5.0 \pm 0.8$ \\
Intrinsic speed (units of $c$) & $\beta$ & $\geqslant 0.980 \pm 0.006$ \\
Bulk Lorentz factor & $\Gamma$ & $\geqslant 5.1 \pm 0.8$ \\
Position angle & $\lambda$ & $31\fdg8 \pm 0\fdg2$\\ 
Inclination angle & $i$ & $\leqslant 11\fdg4 \pm 1\fdg7$ \\ 
Projected half opening angle & $\psi$ & $4\fdg5 \pm 1\fdg3$ \\ 
Intrinsic half opening angle & $\theta_0$ & $\leqslant 0\fdg9 \pm 0\fdg1$ \\ \hline
\end{tabular}}
\caption{Kinematic parameters inferred from the core separation data and the relative position data (Fig. \ref{hoa}).}
\label{tab1}
\end{table}

\begin{table}{
\begin{tabular}{cclcccc}
\hline
Epoch     & $\nu$ & $S_\nu$& $\theta_{\rm maj}$ & Axis  & P.A.        & $\alpha$ \\
          & (GHz) & (mJy)  & (mas)              & ratio & ($^{\circ}$)& \\ \hline \hline
1997.027  & 2.27  & 869    & 1.99           &     0.0     &        21.9     &    0.26\\
          & 8.34  & 617    &     1.01         &         0.038 &        32.7 & \\
1997.344  & 2.27  & 1199   &      1.56          &        0.0  &          29.8 &       0.18\\
          & 8.34  & 943    &       1.04  &        0.315  &       29.5 & \\
1999.466  & 2.29  & 1055   &        1.30 &         0.538&         21.1&         0.26\\ 
          & 8.65  & 750    &       1.06 &          0.287  &        33.6  & \\
2001.823  & 2.30  & 1037   &       1.45  &         0.393  &        34.4  &        0.35\\
          & 8.65  & 652    &       1.61 &          0.157 &         31.2 &  \\
2004.531  & 2.31  & 742    &       2.10  &         0.347 &         30.4 &         0.48\\
          & 8.65  & 395    &       1.92  &         0.143 &         30.1  &    \\ \hline
\end{tabular}}
\caption{Results from the model fitting of $S$ and $X$ band VLBA visibility data. The mean spectral index $\tilde{\alpha} = 0.31$. The fitted $S_\nu$ are assumed to be accurate to 5 \% and the uncertainties for $\theta_{\rm maj}$ are $\leqslant$ 0.01 mas.}
\label{tab2}
\end{table}

\section{Kinematic model}
\label{kinjet}

The kinematic jet models presented in \cite{2015ApJ...805...91M} involve the orbital motion of short-lived flux frozen, relativistic plasma blobs or jet knots (over-dense pockets acted on by centrifugal driving, radiation pressure and gravitational force of the black hole) outflowing along a magnetic flux surface composed of contours of constant magnetic flux, with the field lines anchored to the accretion disk. The underlying bulk plasma flow has a conserved total energy $E$ and angular momentum $L$. A quasi-periodic flux variability (days to months timescale) carrying the signature of the knot helical motion is expected to dominate the jet emission when the instantaneous velocity vector is aligned close to the inclination angle of the jet towards the observer's line of sight. It is an extension of the internal jet rotation scenario proposed by \cite{1992A&A...255...59C} to include general relativistic effects on jet knot kinematics constrained to move along a conical geometry and their observational signatures. The general relativistic model is essential in the context of the knot kinematics and their emission in the vicinity of the black hole which leads to modified observational effects including a gradually additive light curve phase shift, increased amplitude and differing power spectrum slopes when compared to the special relativistic case. The formalism is set in Schwarzschild geometry (outside non-rotating black hole) owing to the ease in the analytic description of gravitational effects including the redshift, light bending (due to curved space-time) and time delay (with respect to a light ray path pointed straight to the observer) which are imprinted on the emitted radiation from the jet knots. In the following description, we ignore light bending and time delay based corrections as they are very small compared to the dominant gravitational redshift. These effects must be accounted for in the light curve and spectra from the inner accretion disk--corona system \cite[e.g.][]{2008MNRAS.384..361D,2015ASInC..12..112M} while we are probing the developing jet at some distance away from this region. 

\subsection{Formalism}
The metric outside a Schwarzschild black hole is given by
\begin{equation}
ds^2 = -(1-2 r_{\rm g}/r) c^2 dt^2+\frac{dr^2}{(1-2 r_{\rm g}/r)}+r^2 (d \theta^2+\sin^2\theta d \phi^2),
\label{metric}
\end{equation}
where $(r,\theta,\phi)$ are the spherical polar coordinates, $r_{\rm g} = G M_{\bullet}/c^2$ is the gravitational radius with black hole mass $M_{\bullet}$ and $t$ is the coordinate time measured by an observer at infinity. The proper time interval as measured by an observer in the instantaneous rest frame (jet frame) is
\begin{equation}
c^2 d\tau^2 = - ds^2 = c^2 dt^2 (1-2 r_{\rm g}/r) \Gamma^{-2},
\label{dtaudt}
\end{equation}
where $\Gamma = (1-\beta^2_\Gamma)^{-1/2}$ is the bulk Lorentz factor with an associated bulk three-velocity $\beta_\Gamma$. The time component of the contravariant four-velocity ${\bf u}$ is
\begin{equation}
u^t = \frac{dt}{d \tau} = \Gamma (1-2 r_{\rm g}/r)^{-1/2}
\label{utGamma}
\end{equation}
The Schwarzschild metric admits two Killing vectors \bm{$\zeta$} and \bm{$\eta$}. In the coordinate system $(t,r,\theta,\phi)$ these vectors have the components $\zeta^{\mu} = (1,0,0,0)$ and $\eta^{\mu} = (0,0,0,1)$ associated with symmetry under changes in $t$ and $\phi$ respectively. The conserved total energy $E$ and angular momentum $L$ of the bulk flow can be expressed in terms of these as 
\begin{align}\label{EL}
E &= -\bm{\zeta}{\bf \cdot u} = (1-2 r_{\rm g}/r)~u^t\\ \nonumber
L &= \bm{\eta}{\bf \cdot u} = r^2 \sin^2 \theta~u^{\phi}.
\end{align}
Using $u^t$ from eqn. (\ref{utGamma}) in the above,
\begin{equation}
E = \Gamma (1-2 r_{\rm g}/r)^{1/2} = \Gamma_F,
\label{Gammaf}
\end{equation}
where $\Gamma_F$ is the asymptotic bulk Lorentz factor at large $r$ such as that measured by an observer at infinity. Thus,
\begin{align}\label{Gammabeta}
\Gamma &= \Gamma_F (1-2 r_{\rm g}/r)^{-1/2}\\ \nonumber
\beta_\Gamma &= (1-\Gamma^{-2})^{1/2} = \left(1-\Gamma_F (1-2 r_{\rm g}/r)\right)^{1/2}.
\end{align}
The four velocity components of the emitting source in the instantaneous rest frame is $u^{\mu}_{\rm rec} = (1,0,0,0)$ and that in the observer frame is $u^{\mu}_{\rm em} = u^t(1,\vec{\bm{\beta}_\Gamma})$. The Doppler factor $D$ is the ratio of the energy of a photon received by an observer at infinity ($E_{\rm rec}$) to that emitted in the instantaneous rest frame ($E_{\rm em}$) and is given by
\begin{equation}
D = \frac{E_{\rm rec}}{E_{\rm em}} = \frac{-{\bf p \cdot u_{\rm rec}}}{-{\bf p \cdot u_{\rm em}}} = \frac{(1-2 r_{\rm g}/r)^{1/2}}{\Gamma (1-\beta_\Gamma \cos \xi)},
\label{Dopplerfactor}
\end{equation}
where $\xi$ is the angle between the photon emission unit vector $\hat{\bm{n}}$ and the velocity vector of the jet knots $\vec{\bm{v}}$. Ignoring the effects of light bending, in the Cartesian coordinate system $(x,y,z)$, $\hat{\bm{n}}$ has the components 
\begin{equation}
n_j = (\sin i,0,\cos i),
\label{nj}
\end{equation}
where $i$ is the inclination angle between the $z-$axis of the jet and the observer line of sight. Then,
\begin{equation}
\cos \xi = \frac{\hat{\bm{n}}\cdot\vec{\bm{v}}}{|\vec{\bm{v}}|}.
\label{cosxi}
\end{equation}
The time interval as measured in the observer frame $d \tilde{\tau}$ is related to that measured by an observer in the instantaneous rest frame $d \tau$ by $d \tilde{\tau} = D^{-1} d \tau$ and hence is related to the coordinate time interval $dt$,
\begin{equation}
\tilde{\tau} = (1+\tilde{z}) \int D^{-1} d \tau = (1+\tilde{z}) \int^{t}_{0} dt (1-\beta_\Gamma \cos \xi)
\label{lambdat}
\end{equation}
The time $\tilde{\tau}$ accounts for the relative motion between the relativistic source and the observer and for cosmological expansion through the redshift factor $\tilde{z}$. From eqn. (\ref{EL}), the azimuthal velocity $\dot{\phi}$ and coordinate $\phi$ are
\begin{align}\label{dotphi}
\dot{\phi} &= \frac{u^\phi}{u^t} = \frac{j (1-2 r_{\rm g}/r)}{r^2 \sin^2 \theta}\\ \nonumber
\phi &= \int^{t}_{0} dt \frac{j (1-2 r_{\rm g}/r)}{r^2 \sin^2 \theta},
\end{align}
where $j = L/E$ is the specific angular momentum. Using the known $\tilde{\tau} = \tilde{\tau}(t)$, $r$ and the velocity components $v^i$ can be expressed in terms of $\tilde{\tau}(t)$. The flux density $S_{\nu,0}$ of photons emitted in the instantaneous rest frame is subjected to Doppler beaming and is received by the observer at infinity at a flux density $S_\nu$. These are related by
\begin{equation}
S_\nu = D^{3+\alpha} S_{\nu,0},
\label{flux}
\end{equation}
where $\alpha$ is the spectral index in the relation $S_\nu \propto \nu^{-\alpha}$ in the observer's frame. For a beamed source, the variable flux density due to relativistic jet knots can be studied by determining $D = D(\tilde{\tau})$ and hence $S_\nu = S_\nu(\tilde{\tau})$ as measured in the observer's frame which on comparison with the observed variable flux density can be used to infer the presence of helical motion. 


\subsection{Component kinematics and emission}
\label{compkin}

At the Alfv\'{e}n radius $\varpi_A$, the constrained bulk flow is in co-rotation with the anchored magnetic flux surface. Beyond $\varpi_A$, but within the light cylinder radius $\varpi_L$, inertial effects due to the co-rotating surface dominate and the bulk flow acquires the asymptotic angular momentum $L$. The Keplerian angular velocity of this co-rotating flux surface is
\begin{equation}
\Omega_f = \varpi^{-3/2}_f c r^{1/2}_{\rm g},
\end{equation}
at the foot point radius $\varpi_f = 6 r_g$ (radius at which the magnetic field lines are anchored to the rotating accretion disk, taken to be the innermost stable circular orbit, ISCO), and the corresponding light cylinder radius is
\begin{equation}
\varpi_L = \Omega^{-1}_f c = \varpi^{3/2}_f r^{-1/2}_{\rm g}.
\end{equation}
The launch radius of the jet knots is
\begin{equation}
\varpi_0 = f \varpi_L,
\label{varpi0}
\end{equation}
where the scaled launch radius $f \leq 10$ \cite[e.g.][]{1992A&A...255...59C,1997A&A...319.1025F}. As $\varpi_0 > \varpi_f$, $f > \varpi_f/\varpi_L$ and hence, $f > \varpi^{-1/2}_f r^{1/2}_{\rm g}$. The Alfv\'{e}n radius is
\begin{equation}
\varpi_A = x_A \varpi_L,
\end{equation}
where $x^2_A \leq 1$.  As $\varpi_f = 6 r_{\rm g} < \varpi_A \leq \varpi_L$, $\varpi_f/\varpi_L < x_A \leq 1$ and hence, $\varpi^{-1/2}_f r^{1/2}_{\rm g} < x_A \leq 1$. For a conical jet, the knot position $\vec{x}_s$ and velocity $\vec{v}$ are expressed in terms of the cylindrical radius $\varpi$, azimuthal angle $\phi$ and the $z$ distance along the jet axis along with their associated velocities $\dot{\varpi}$, $\phi$ and $\dot{z}$ respectively. In the Cartesian coordinate system $(x,y,z)$, their components are
\begin{align}\label{xs}
x^j_s &= (\varpi \cos \phi, \varpi \sin \phi, z),\\ \nonumber
v^j &= (\dot{\varpi} \cos \phi-\varpi \dot{\phi} \sin \phi, \dot{\varpi} \sin \phi+\varpi \dot{\phi} \cos \phi, \dot{z}).
\end{align}
The cylindrical radius $\varpi$ and the velocity $\dot{\varpi}$ are
\begin{align}\label{varpi}
\varpi &= \varpi_0 + z \tan \theta_0, \\ \nonumber
\dot{\varpi} &= \dot{z} \tan \theta_0.
\end{align}
The radial distance $r$ and the velocity $\dot{r}$, using eqns. (\ref{xs}) and (\ref{varpi}) are
\begin{align}\label{rdotr}
r &= (x^2+y^2+z^2)^{1/2} = \cot \theta_0 (2 \varpi (\varpi-\varpi_0)+\varpi^2_0)^{1/2},\\ \nonumber
\dot{r} &= \frac{\dot{z}}{r} (\varpi \tan \theta_0+z). 
\end{align}
For $\sin \theta = \varpi/r$, the polar velocity $r \dot{\theta}$, using eqns. (\ref{xs}), (\ref{varpi}) and (\ref{rdotr}) is
\begin{equation}
r \dot{\theta} = -\frac{\dot{z} \varpi_0}{r}.
\label{rdottheta}
\end{equation}
As $\varpi=r \sin \theta$, and using eqns. (\ref{utGamma}) and (\ref{EL}), the conserved specific angular momentum $j$, azimuthal velocity $\dot{\phi}$ and coordinate $\phi$ are
\begin{align}\label{dotphiphi}
j &= \varpi^2 (1-2 r_{\rm g}/r)^{-1} \dot{\phi} = \varpi^2_A (1-2 r_{\rm g}/\varpi_0)^{-1} \Omega_f,\\ \nonumber
\dot{\phi} &= \frac{j (1-2 r_{\rm g}/r)}{\varpi^2},\\ \nonumber
\phi &= \int^{t}_0 dt  \frac{j (1-2 r_{\rm g}/r)}{\varpi^2}.
\end{align}
From the condition $u^{\beta} u_\beta = -c^2$ and using $\dot{r}$, $r^2 \dot{\theta}^2$, $\dot{\phi}^2$ from eqns. (\ref{rdotr}), (\ref{rdottheta}) and (\ref{dotphiphi}) respectively, with the Lorentz factor from eqn. (\ref{Gammaf}), the velocity along the jet axis $\dot{z}$ is
\begin{equation}
\frac{\dot{z}}{c} = r (1-2 r_{\rm g}/r) \left(\frac{1-(1-2 R_{\rm g}/r) \left(\Gamma^{-2}_F+\displaystyle{\frac{j^2}{\varpi^2 c^2}}\right)}{(\varpi \tan \theta_0+z)^2+(1-2 r_{\rm g}/r) \varpi^2_0}\right)^{1/2}.
\end{equation}
The above equation is integrated to give $z = z(t)$ and using the relation between $\tilde{\tau}$ and $t$ from eqn. (\ref{lambdat}), we obtain $z = z(\tilde{\tau})$. Using this, we can evaluate $\varpi (\tilde{\tau})$, $\dot{\varpi} (\tilde{\tau})$, $\phi (\tilde{\tau})$ and $\dot{\phi} (\tilde{\tau})$ and the changing angle $\cos \xi (\tilde{\tau})$ from eqn. (\ref{cosxi}), which when expressed in terms of $\vec{n}$ from eqn. (\ref{nj}) and $\vec{v}$ from eqn. (\ref{xs}) is 
\begin{equation}
\cos \xi = \frac{\dot{\varpi} \cos \phi \sin i-\varpi \dot{\phi} \sin \phi \sin i+\dot{z} \cos i}{(\dot{\varpi}^2+\varpi^2 \dot{\phi}^2+\dot{z}^2)^{1/2}}
\end{equation}
The Doppler factor in eqn. (\ref{Dopplerfactor}) is then $D = D(\tilde{\tau})$ and hence, the variable flux density in eqn. (\ref{flux}) is $S_\nu = S_\nu (\tilde{\tau})$.

In order to infer the helical motion of jet knots in the observer's frame, the instantaneous jet positions $x^{j}_s$ from eqn. (\ref{xs}) are transformed to the observer's Cartesian coordinates $x^{j}_{\circ}$ using $x^{j}_\circ = {\bf R}_\lambda {\bf R}_i x^{j}_s$ where ${\bf R}_\lambda$ and ${\bf R}_i$ are Euler rotation matrices about the observer's $Z$-axis and $Y$-axis respectively \cite[e.g. as used in][]{2014MNRAS.445.1370K}, using the inclination angle $i$ and a translated position angle $\tilde{\lambda} = \pi/2-\lambda$. Then,
\begin{align}\label{x0}
x_\circ &= \varpi \cos \phi \cos i \cos \tilde{\lambda}+z \sin i \cos \tilde{\lambda}-\varpi \sin \phi \sin \tilde{\lambda},\\ \nonumber
y_\circ &= \varpi \cos \phi \cos i \sin \tilde{\lambda}+z \sin i \sin \tilde{\lambda}-\varpi \sin \phi \tilde{\cos \lambda},\\ \nonumber
z_\circ &= -\varpi \cos \phi \sin i+z \cos i.
\end{align} 
With the instantaneous coordinates $x^j_s = x^j_s (\tilde{\tau})$, we determine $x^{j}_{\circ} = x^{j}_{\circ} (\tilde{\tau})$ to compare the model with the observed component positions where $x_\circ =$ relative R.A. and $y_\circ =$ relative DEC. 

\section{Model application and discussion}
\label{Modelapp}

\begin{figure}
\centering
\includegraphics[scale=0.23]{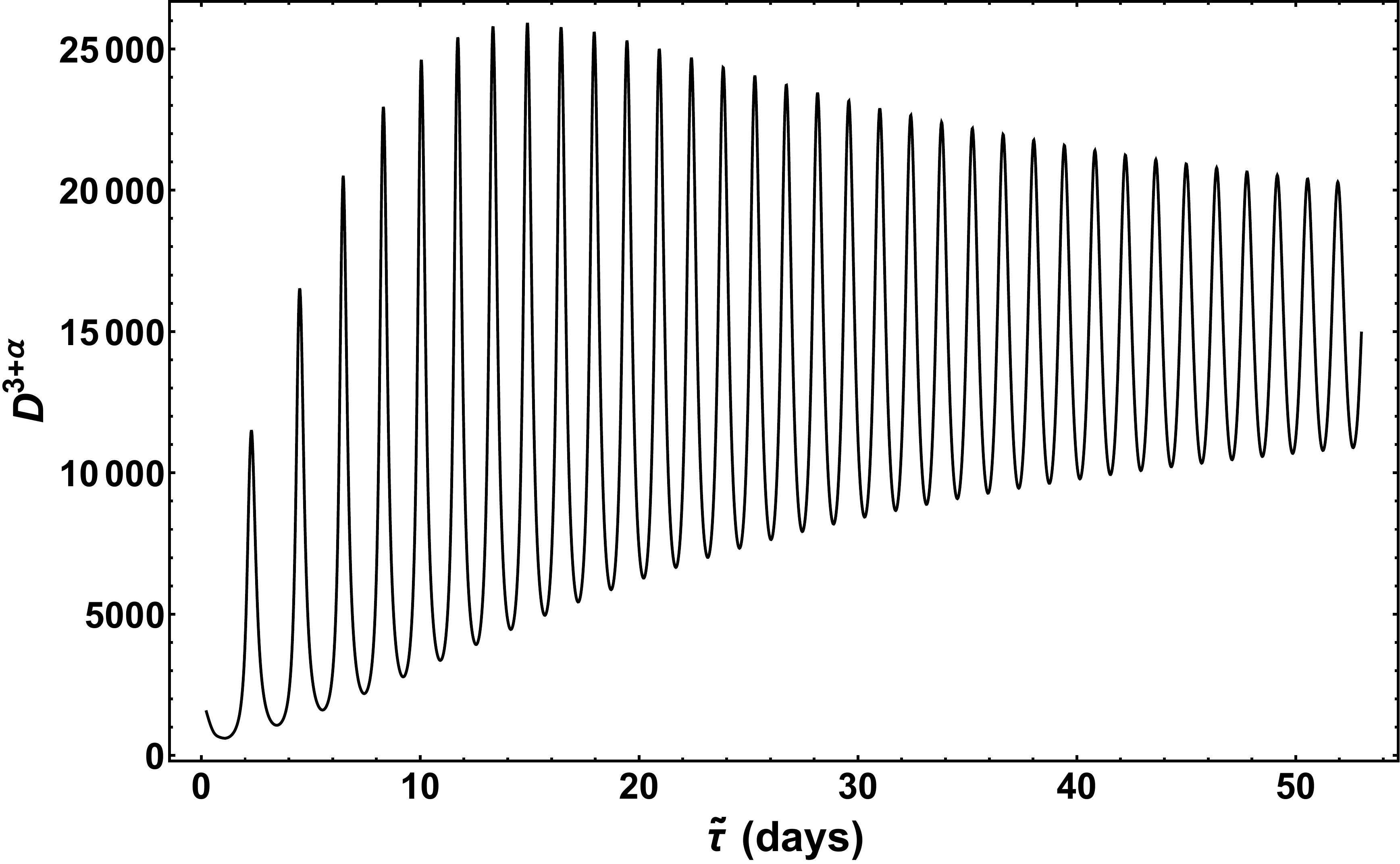}\\
\includegraphics[scale=0.2]{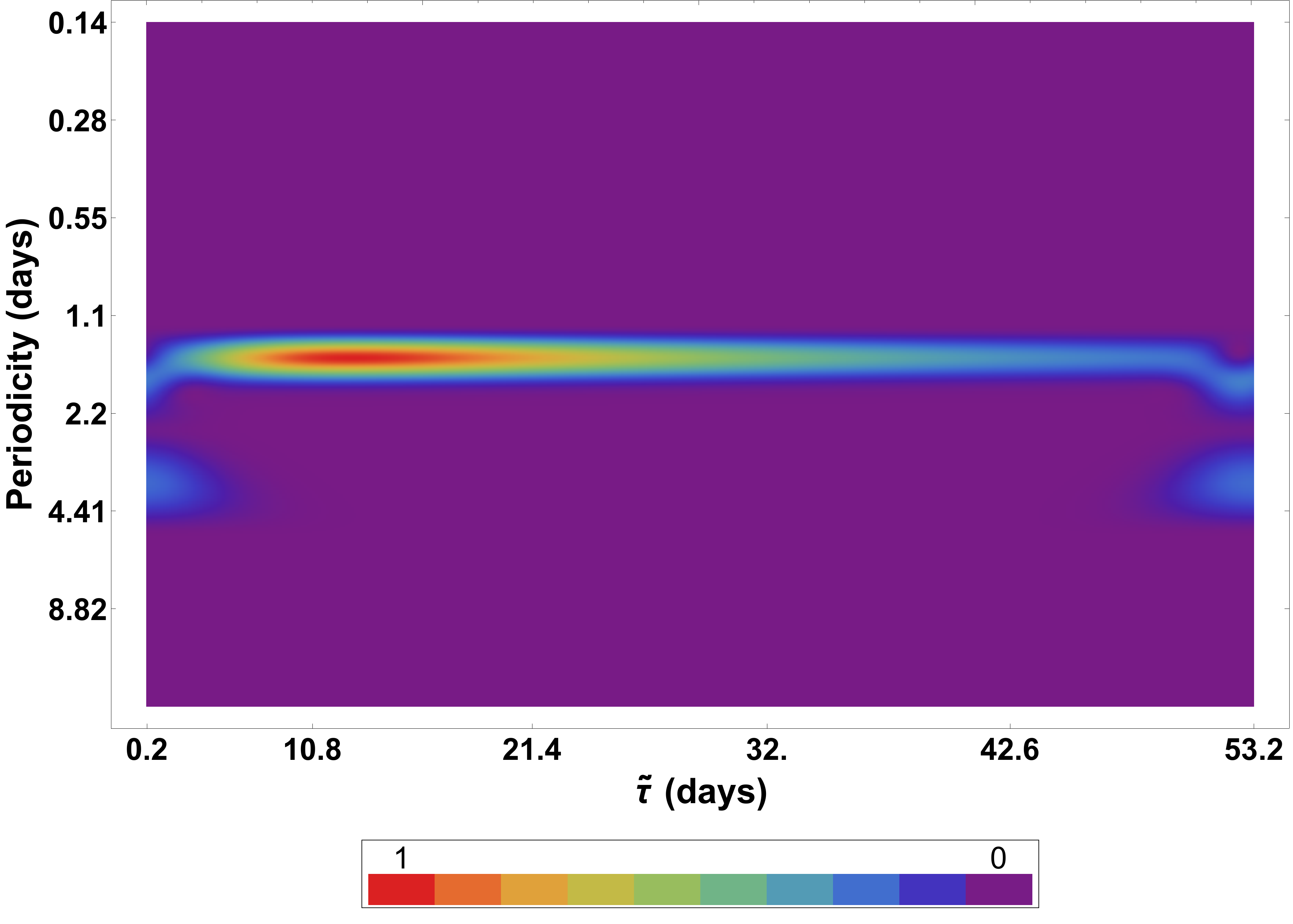}\\
\includegraphics[scale=0.2]{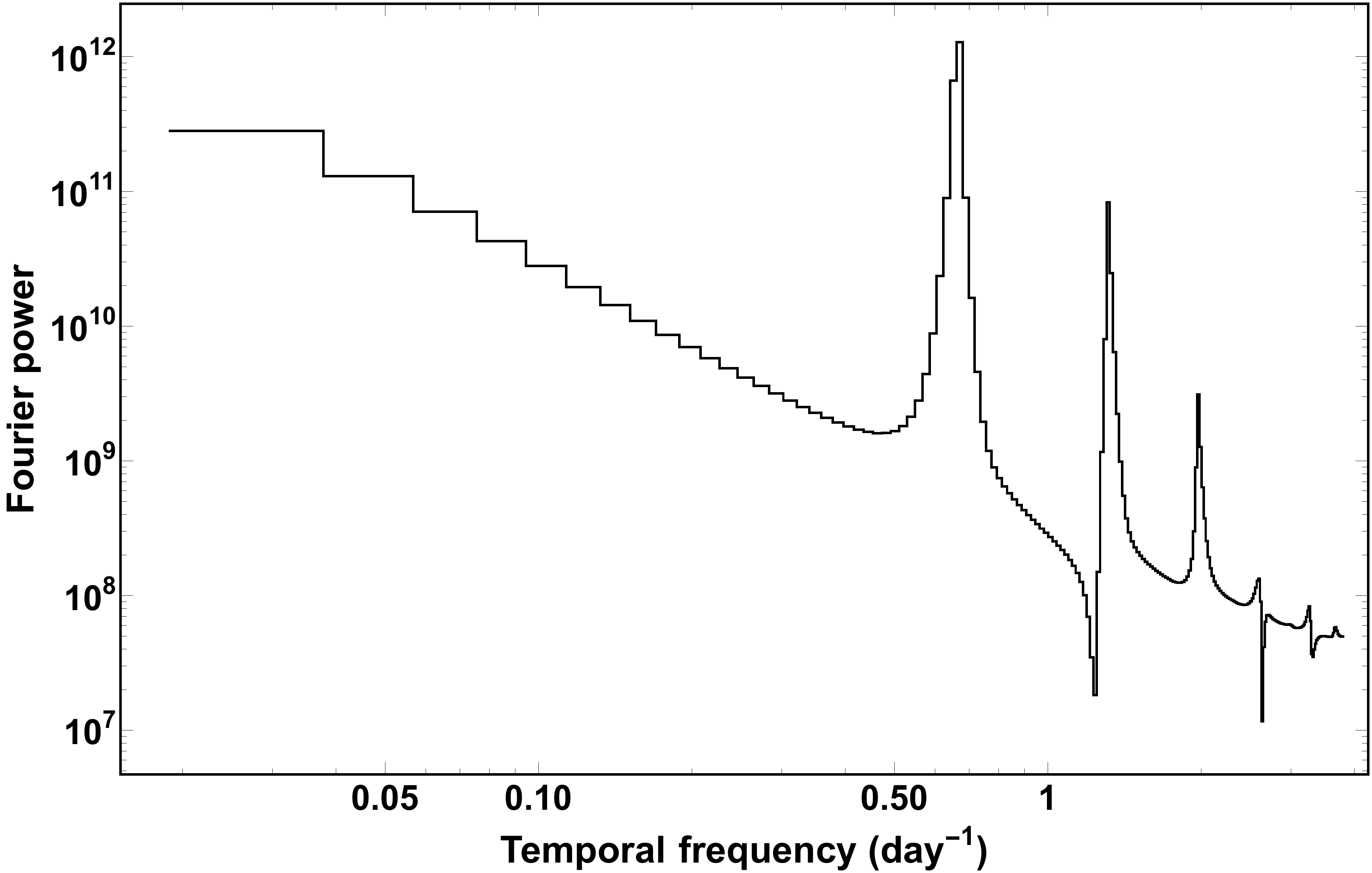}
\caption{With parameters $f = 10$, $\alpha = 0.31$, $\Gamma = 10.8$, $\theta_0 = 0\fdg2$ and $i = 2\fdg2$; Top: simulated light curve showing QPO evolution; Middle: wavelet power spectrum showing the persistent QPO at $1.5$ days with strong power concentration (red portion) during the beaming activity; Bottom: Fourier power spectrum showing evolving QPO centered at $1.6$ days, its weaker harmonics and indicating a power law shape with index $-2.3$.}
\label{simlc1}
\end{figure}

\begin{figure}
\centering
\includegraphics[scale=0.23]{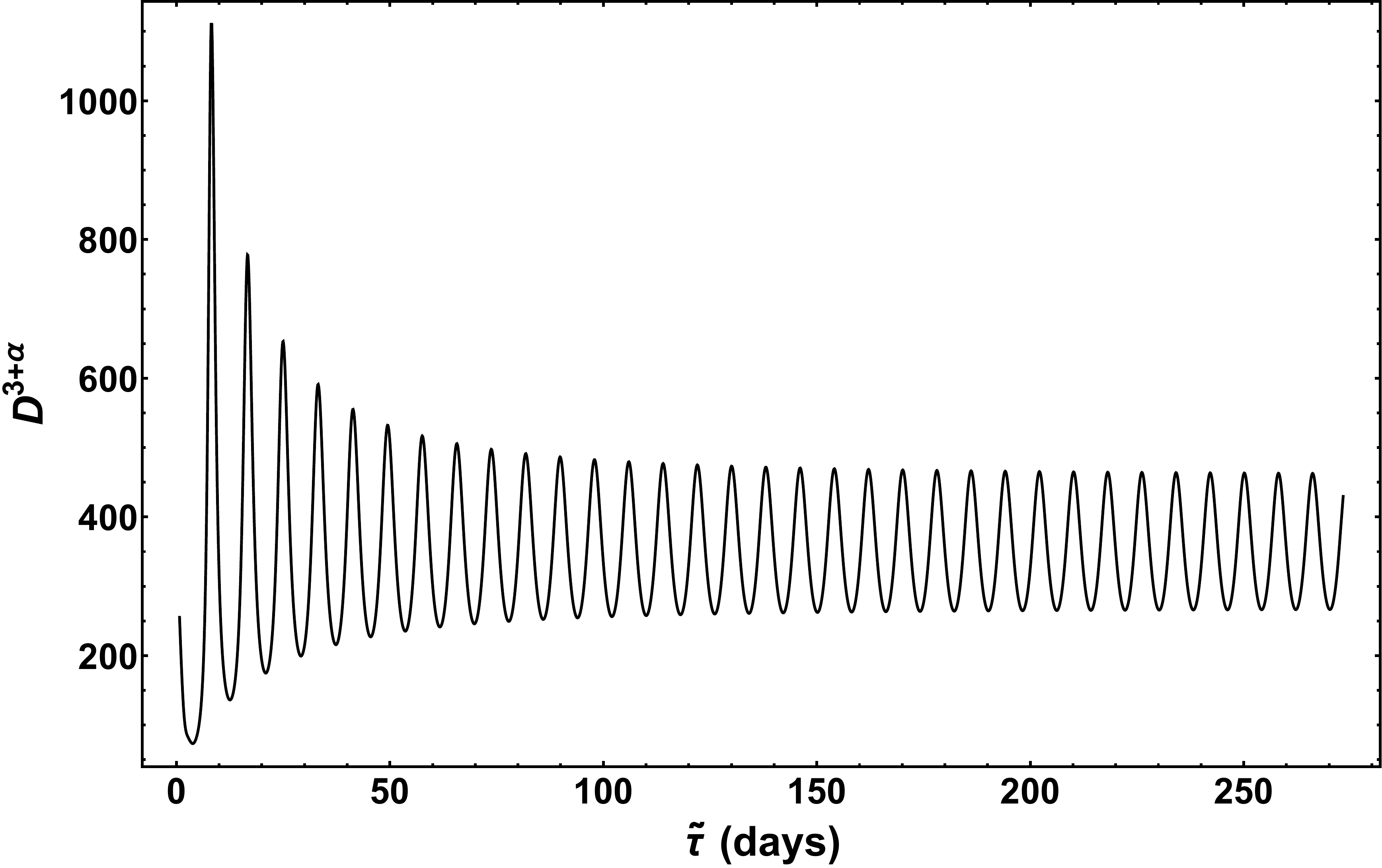}\\
\includegraphics[scale=0.2]{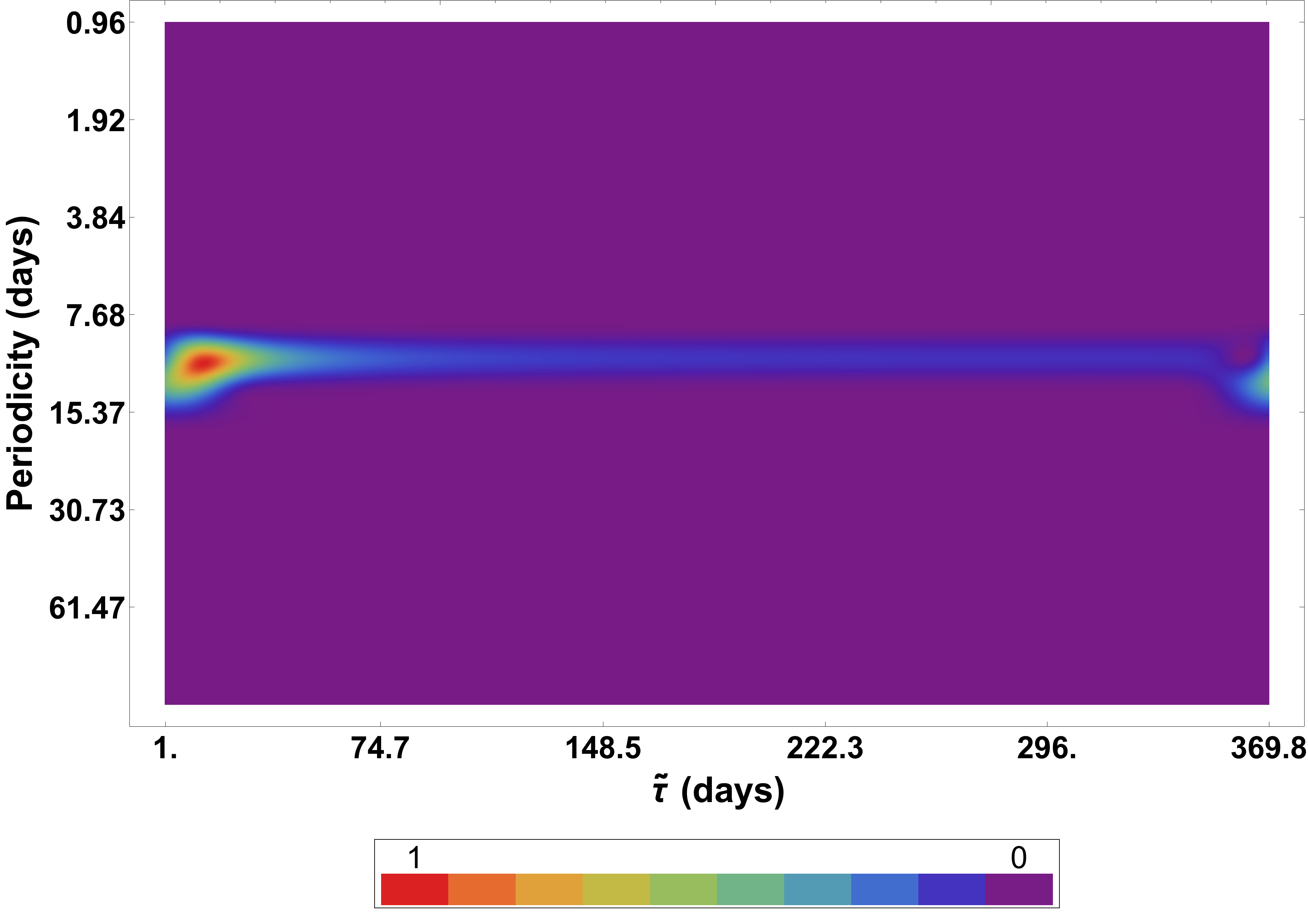}\\
\includegraphics[scale=0.2]{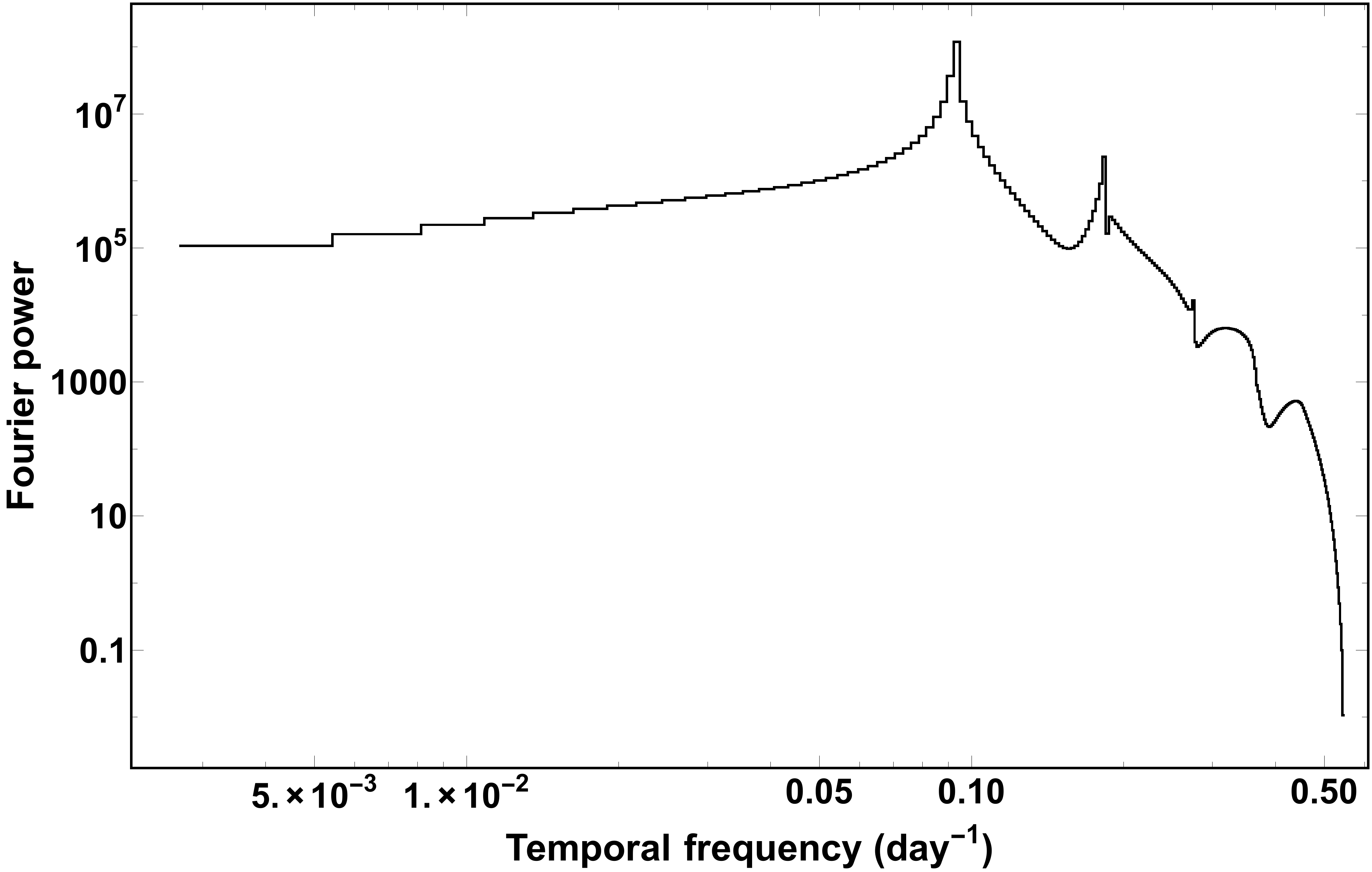}
\caption{With parameters $f = 10$, $\alpha = 0.31$, $\Gamma = 5.1$, $\theta_0 = 0\fdg9$ and $i = 11\fdg4$; Top: simulated light curve showing QPO evolution; Middle: wavelet power spectrum showing the persistent QPO at $10.5$ days with strong power concentration (red portion) during the beaming activity;  Bottom: Fourier power spectrum showing evolving QPO centered at $10.5$ days, its weaker harmonics and indicating a flattening of the power spectrum at lower temporal frequencies.}
\label{simlc2}
\end{figure}

For PG 1302$-$102 with $M_\bullet = 4 \times 10^8 M_\odot$, the ratio $r_{\rm g}/\varpi_0 \geq 0.007$ for the knot launch radius $\varpi_0 \leq 10 \varpi_L \sim 147 r_{\rm g}$ indicating that relativistic gravitational effects will be prominent for emission from knots launched close to the black hole. To illustrate these effects, we simulate the expected variable light curve due to beamed emission from jet knot motion for $f = 10$ in two cases (i) $\Gamma = 10.8$, $\theta_0 = 0\fdg2$ and $i = 2\fdg2$ obtained in \cite{2015MNRAS.454.1290K}, corresponding to the extremal case and (ii) the equalities in the limits $\Gamma \geq 5.1$, $\theta_0 \leq 0\fdg9$ and $i \leq 11\fdg4$ from the current study, corresponding to the mean jet properties. The simulated light curves showing the QPO along with their associated wavelet power spectrum \cite[distribution of spectral power as a function of period and $\tilde{\tau}$, e.g.][and references therein]{2016MNRAS.456..654M} and Fourier power spectrum are presented in Fig. \ref{simlc1} for case (i) and Fig. \ref{simlc2} for case (ii). 

The QPO for case (i) is centered at $1.6$ days and for case (ii) at $10.5$ days. Both are characterized by QPO timescales $\sim$ 10 days as expected in this scenario for massive quasars \cite[e.g.][]{2004ApJ...615L...5R}, and the presence of weak harmonics and high frequency power law spectral shape. 
The spectral shapes obtained in the above simulations are consistent with observationally inferred shapes in a wide variety of AGN \cite[e.g.][and references therein]{2014ApJ...791...74M,2015MNRAS.452.2004M,2016MNRAS.456..654M} for which the power law index typically ranges between $-1.0$ and $-2.5$. The wavelet power spectrum of the simulated light curves shows that the QPO is persistent throughout the duration, with power concentrated in portions during the beaming activity (red portions in the wavelet power spectrum), the transient phase during which the QPO can be inferred. The variable emission originates from the disk-jet region consisting of the inner accretion disk, a surrounding corona and the developing jet.

Multiwavelength flux density variability in radio-loud AGN spans a diverse range of timescales from hours to months and even years \cite[e.g. as discussed in][and references therein]{2016MNRAS.456..654M}. Aperiodic variability accompanied by strong flaring is mainly contributed to by turbulent cells in the jet which can give rise to the observed beamed emission at random intervals due to the relative motion and interaction with shocks \cite[e.g.][]{1979ApJ...232...34B,2008Natur.452..966M,2014ApJ...780...87M}, with a spectral energy distribution contributed to by synchrotron and inverse Compton emission \cite[e.g.][]{1981ApJ...243..700K}. Quasi-periodic variability associated with helical jet motion can be produced by multiple mechanisms. Intrinsic jet-based mechanisms include the use of angular momentum and accretion energy supplied by the accretion disk for an internally rotating jet constituting a bulk plasma flow \cite[][]{1992A&A...255...59C}. External mechanisms include changes in the accretion flow, which can also be externally driven by a SMBBH, causing a warped accretion disk due to torquing action \cite[e.g.][]{2004ApJ...602..625C,2004ApJ...615L...5R,2013ApJ...765L...7N}; the SMBBH--accretion disk tidal interaction which leads to the setting up of shocks in the disk and strong fluctuations in the mass accretion rate resulting in periodic flaring such as the 12-yr periodic double-peaked outbursts in OJ 287 \cite[e.g.][]{1988ApJ...325..628S,1996ApJ...460..207L,2008Natur.452..851V}; jet precession due to a closely separated SMBBH causing a warped inner disk or thick disk oscillations which is accompanied by quasi-periodic injection of particles into the jet producing quasi-periodic timescales and their harmonics spanning months--years \cite[e.g.][]{2006ApJ...650..749L,2013MNRAS.434.3487A} or the development of instabilities such as the Kelvin--Helmholtz instability \cite[e.g.][and references therein]{2010MNRAS.402...87A,2011A&A...529A.113Z,2014MNRAS.443...58W}. 

The Keplerian timescale $P$ for a spinning black hole (Kerr space-time outside the black hole) with dimensionless spin parameter $a$ is given by 
\begin{equation}
P = 3.1 \times 10^3 (1+\tilde{z}) \left\{ \left(\frac{r}{r_{\rm g}}\right)^{3/2}+a\right\} \left(\frac{M_\bullet}{10^8 M_\odot}\right)  {\rm s}.
\end{equation}
In the intrinsic jet-based scenarios, with $r = r_{\rm ISCO} = 6~r_{\rm g}$ (non-rotating black hole with $a=0$), $P = 2.3 \times 10^5$ s ($2.7$ days) and with $r = r_{\rm ISCO} = 1.24~r_{\rm g}$ (maximally spinning black hole with $a = 0.998$), $P = 3.8 \times 10^4$ s ($0.4$ days), indicating the typical timescale consistent with the inference from the simulated light curves. In the binary scenarios (with a separation $d$), $r = (0.003 - 0.01)$ pc \cite[][]{2015Natur.518...74G,2015Natur.525..351D}, and hence $P = (3.1 \times 10^7 - 1.9 \times 10^8)$ s ($1 - 6$ yrs). As there is at least 2 orders of magnitude difference between these scenarios, one can distinguish between the identification of both in observational data. 

It must be noted that as $r_{\rm ISCO} = (1.24 - 6) r_{\rm g}$ ($a = 0.998$ and $a = 0$ respectively), the foot point angular velocity $\Omega_f$ of the anchored magnetic flux surface is larger, leading to differences in subsequently calculated quantities in Section \ref{compkin}. In addition, the most general problem involves the self-consistent magnetic flux generation and stability over large scales (here assumed to be conically structured for simplicity) and the subsequent orbital motion of jet knots in this realistic geometrical setup. This requires solving the relativistic Grad--Shafranov equation in Kerr space-time to determine the magnetic flux evolution and time dependence and eventual structuring of the bulk flow and jet knot kinematics, which is beyond the scope of the current work as our focus here is mainly on the knot kinematics and expected emission due to a fixed observationally relevant conical jet shape and its differentiation from observational signatures in a binary black hole scenario.
We propose to include a detailed self-consistent model of the field geometry in Kerr background in the future to calculate the kinematics. In simple models where an in situ generation of the magnetic field is not considered, the magnetic field should satisfy the generalized relativistic Grad--Shafranov equation that includes plasma pressure, which has been addressed in previous works \cite[e.g.][]{1997PhyU...40..659B,2001A&A...365..631F,2001A&A...369..308F}. The special case of a force-free configuration has been considered in several works \cite[e.g.][]{2005ApJ...620..889U}, where the magnetosphere around a Kerr black hole connecting it to a thin accretion disk is numerically evolved to investigate parameter regimes that govern the development of closed and open field geometries, the black hole spin controlling the radial extent of the closed force-free link to the disk surface.
The use of a funnel shaped jet approximating this realistic scenario by \cite{2015ApJ...805...91M} leads to slightly differing results from the conical jet scenario in terms of the light curve phase shift and amplitude, though giving similar QPO timescales and thus being consistent with the results discussed above.

\begin{figure}
\centerline{\includegraphics[scale=0.25]{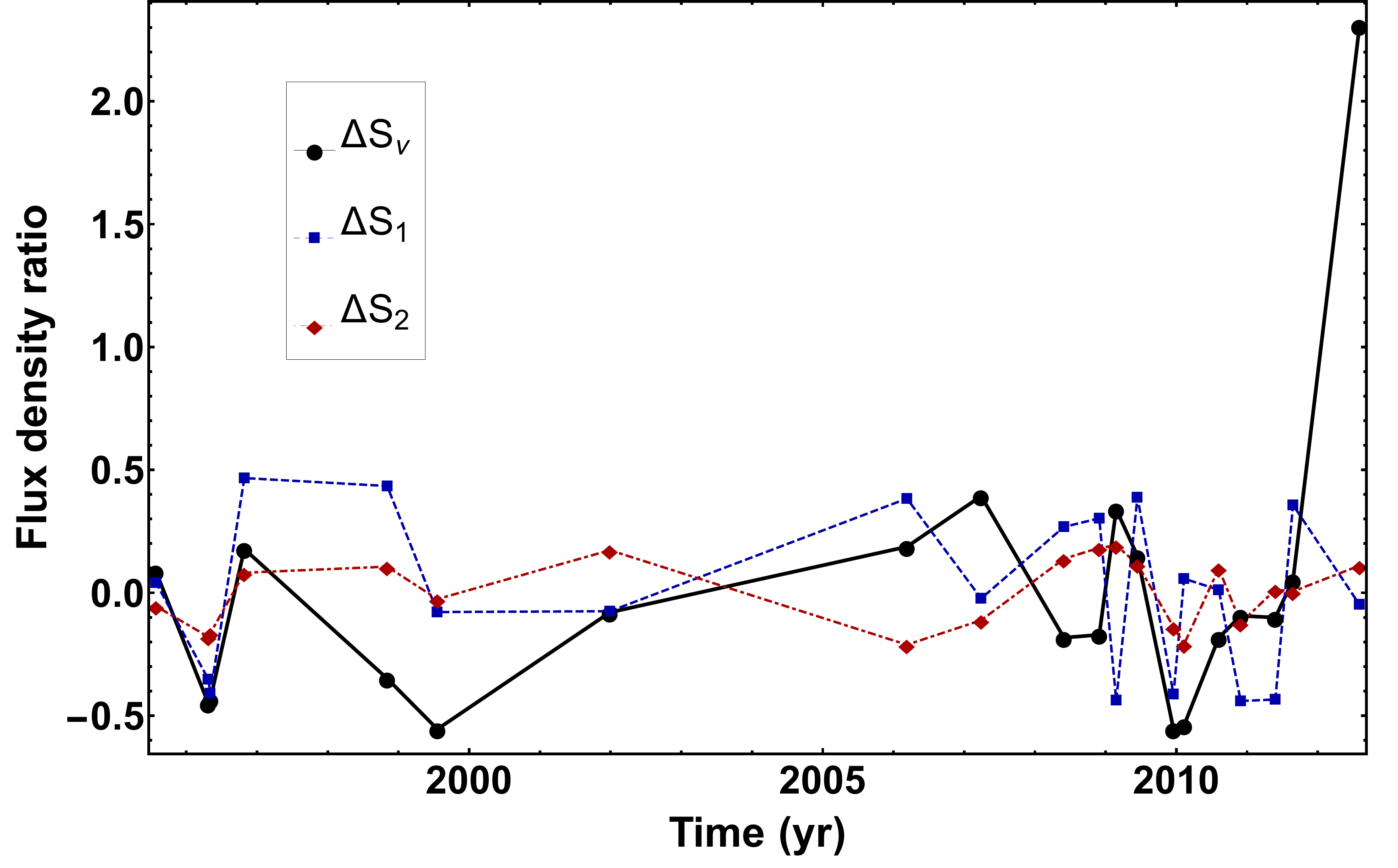}}
\caption{Comparison of the observed core flux density variation (thick black line) with expectation from sinusoidal light curves with periodicities 1.6 days (blue dashed line; case i) and 10.5 days (red dot-dashed line; case ii).}
\label{coremodel}
\end{figure}

The absence of a long term and the day scale QPOs in the current data (see Fig. \ref{Snuobs}) is likely due to the sparse time sampling of the MOJAVE observations. The observed component flux density variability is more likely dominated by propagating shocks in the jet or turbulent processes \cite[e.g.][]{1979ApJ...232...34B,1985ApJ...298..114M,2014ApJ...780...87M}. However, as the core flux density variations can contain the Doppler boosted signature of an underlying helical jet, the flux density variations as obtained for cases (i) and (ii) are compared with observations. From the simulated light curves for cases (i) and (ii), we first evaluate $D_{R,j} = D^{3+\alpha}_j/\bar{D}^{3+\alpha}_j$ where $\bar{D}^{3+\alpha}_j$ is the mean of the simulated light curve $D^{3+\alpha}_j$ with $j=1,2$. Then, we generate two sinusoidal light curves, $S_j$ with period corresponding to the QPO central values (corresponding to the beamed portion), amplitude equal to the standard deviation of $D_{R,j}$, and mean equal to the mean of $D_{R,j}$. Then, we sample $S_j$ at the same times as the observed core flux density observations. Finally, for each $S_j$, we determine $\Delta S_{j} = (S_{j}-\bar{S}_{j})/\bar{S}_{j}$ where $\bar{S}_{j}$ is the mean of $S_{j}$, and for the observed flux density, we determine $\Delta S_\nu = (S_\nu-\bar{S_\nu})/\bar{S}_\nu$ where $\bar{S}_\nu$ is the mean of the core flux density variation. The comparison between the data and the expectation from the simulations are presented in Fig. \ref{coremodel}. 

The contribution of a helical knot based flux density change to the observed core flux density is quantified using $\sigma_{j}/\sigma_{\nu}$ where $\sigma_{j}$ and $\sigma_{\nu}$ are the standard deviations of $\Delta S_j$ and $\Delta S_\nu$ respectively, the ratio ranging between $0.23 - 0.53$ indicating that signatures from the beamed emission are present in the core flux density variation. The contribution can be further ascertained with observations of spectral variability \cite[e.g.][]{2004A&A...419..913O}, and systematic gradients in the rotation measure from polarization measurements at the parsec scales \cite[e.g.][]{2009MNRAS.393..429O,2012AJ....144..105H}, also expected in simulations \cite[e.g.][]{2010ApJ...725..750B,2016ApJ...817...63Z} which can offer evidence for a helical jet dominated by magnetic fields.

\begin{figure}
\centering
\includegraphics[scale=0.24]{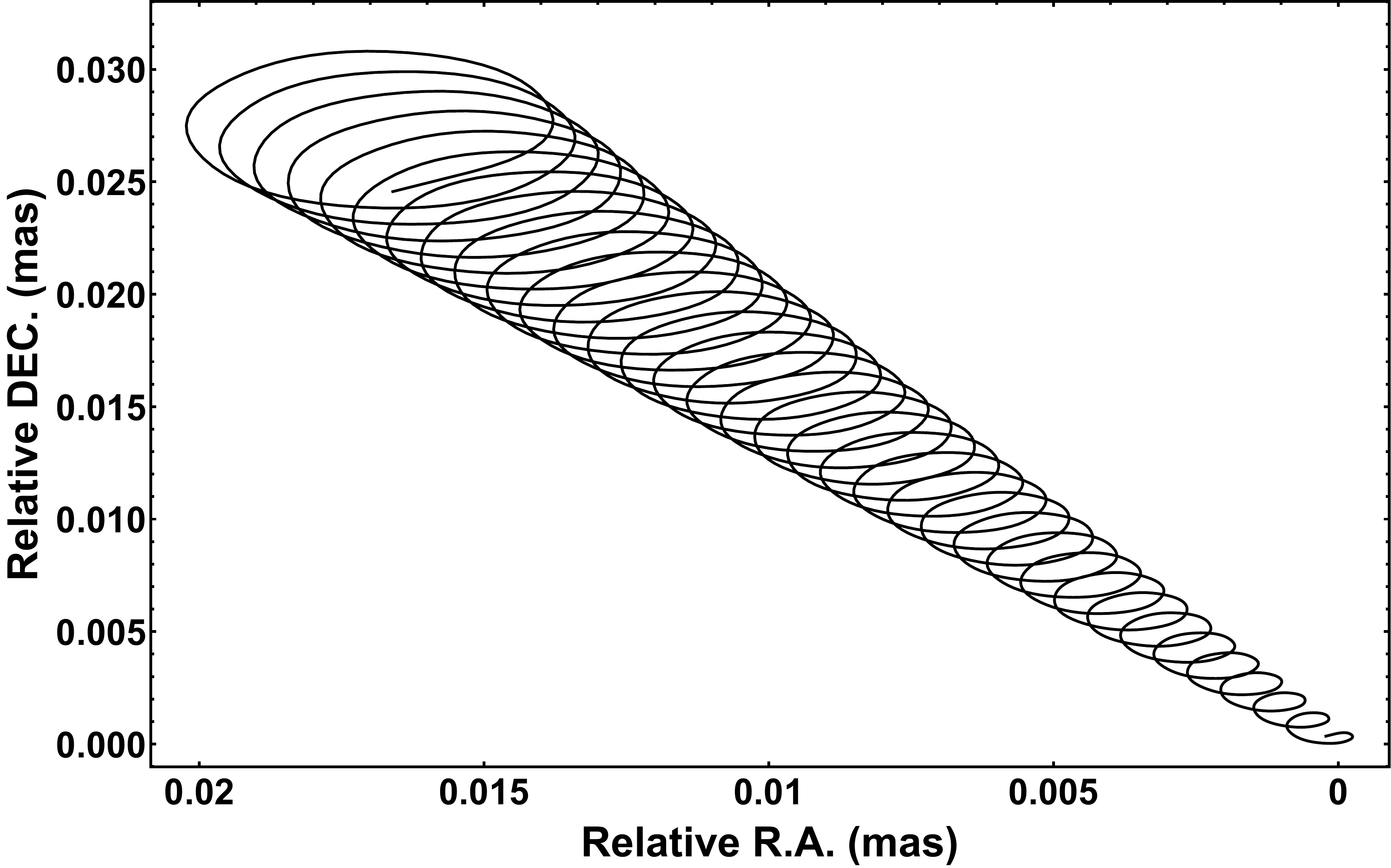}\\
\includegraphics[scale=0.24]{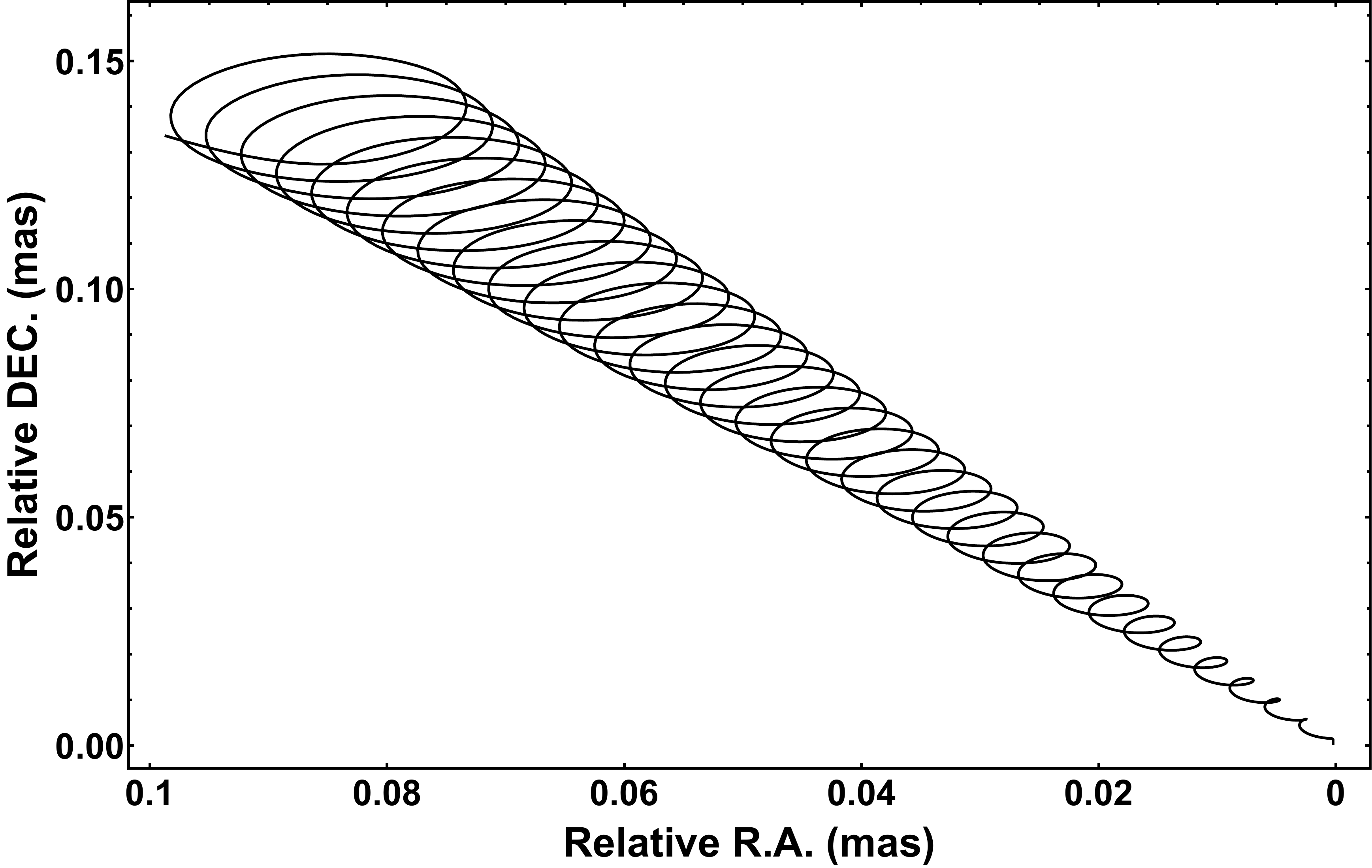}
\caption{Expected projected trajectory of components launched from $f = 10$ and $\lambda = 31\fdg8$. Top: parameters combination of $\Gamma = 10.8$, $\theta_0 = 0\fdg2$ and $i = 2\fdg2$; Bottom: parameters combination of $\Gamma = 5.1$, $\theta_0 = 0\fdg9$ and $i = 11\fdg4$. The time span for each trajectory is the same as that for the simulated light curves in Figs. \ref{simlc1} and \ref{simlc2}.}
\label{xyplot}
\end{figure}

The expected helical trajectory of a knot is evaluated using eqn. (\ref{x0}) for the two cases of parameter combinations used earlier (cases i and ii) and a position angle $\lambda = 31\fdg8$. The time taken for the trajectory to develop is the same as that evaluated for the simulated light curves in Figs. \ref{simlc1} and \ref{simlc2} and covers the beamed duration. The trajectories are plotted in Fig. \ref{xyplot}. 

\begin{figure}
\centering
\includegraphics[scale=0.25]{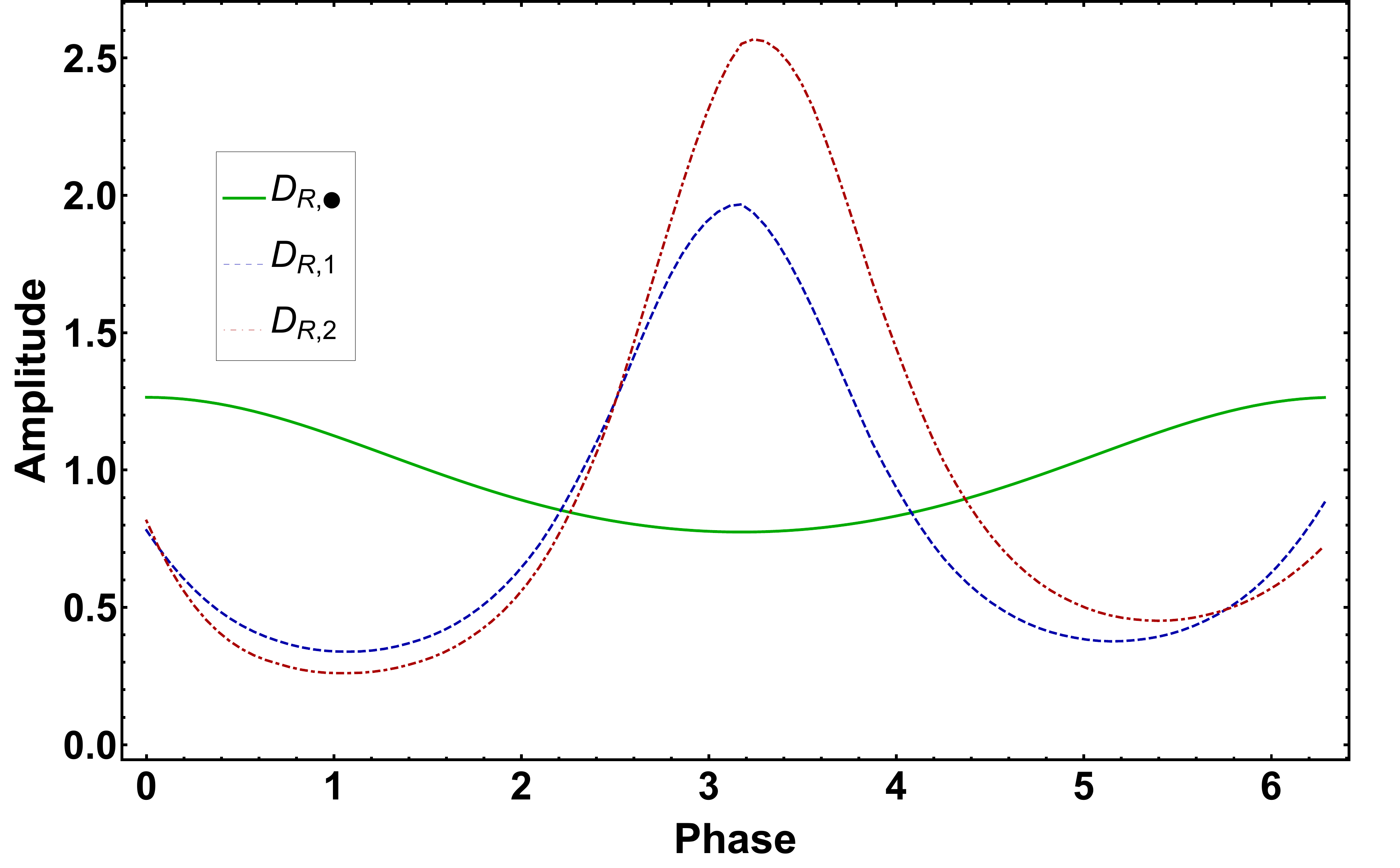}
\caption{Comparison of $D_{R,\bullet}$ (thick green line) with $D_{R,j}$ (blue dashed line for $j=1$ and red dot-dashed line for $j=2$). There is a noticeable phase difference $\sim \pi$ between them and the amplitude ratio $D_{R,j}/D_{R,\bullet} = 0.2 - 3.3$ indicating that flux density variations due to helical motion (transient and short lived) will be dominant over short observation intervals (e.g. days--months).}
\label{DRplot}
\end{figure}

A careful monitoring using high resolution observations of the core and the jet base over a duration spanning a few years with a moderate sampling will be able to identify the observational signatures including flux density variability and a helical trajectory. For knots launched at $\varpi_0 = 10 \varpi_L \sim 0.003$ pc, the ratio $\varpi_0/d = 0.3 - 1$ implying that binary activity will strongly influence flux density variability and knot motion. Any systematic departure from the calculated trajectory (inclination angle and position angle drift) and flux density variability amplitudes as expected for the SMBBH scenario \cite[][]{2014MNRAS.445.1370K} due to accumulative perturbation effects which will strongly indicate the presence of a binary black hole. Optical surveys such as the CRTS, based on the identification of $\sim$ years timescale QPOs from AGN light curves \cite[][]{2015MNRAS.453.1562G} offer a promising method of detecting SMBBH systems. However, the study of \cite{2016MNRAS.461.3145V} finds that AGN stochastic variability, characterized by a red noise power spectrum shape can result in spurious detections owing to observational sampling, necessitating the calibration of the false positive rate by accounting for large source samples dominated by stochastic variability. There is thus a strong requirement for high resolution monitoring observations of possible SMBBH candidates through direct imaging. In addition, the expected flux density variation from the Doppler beaming of the binary orbital signature can be used for this inference. The variable Doppler factor due to binary orbital motion is \cite[e.g.][]{2015Natur.525..351D}
\begin{equation}
D_{\bullet} = \frac{1}{\Gamma_{\bullet} (1-\beta_{\bullet} \cos \tilde{\phi} \sin \tilde{i})},
\end{equation}
where $\beta_\bullet$ is the orbital velocity of the SMBBH scaled in units of $c$, $\Gamma_\bullet = (1-\beta^2_\bullet)^{-1/2}$, and $\tilde{\phi}$ and $\tilde{i}$ are the orbital phase and orbital inclination, respectively. The corresponding normalized flux density variation is then $D_{R,\bullet} = D^{3+\alpha}_\bullet/\bar{D}^{3+\alpha}_\bullet$ where $\bar{D}^{3+\alpha}_\bullet$ is the mean of $D^{3+\alpha}_\bullet$. Using $\alpha = 0.31$ and $\beta_\bullet \sin \tilde{i} = 0.074$ \cite[][]{2015Natur.525..351D}, $D_{R,\bullet}$ corresponding to a complete binary orbital cycle (periodicity in years timescale) is compared with $D_{R,j}$ corresponding to a complete jet knot orbital cycle obtained during the maximum beaming phase (periodicity in days timescale), to illustrate the relative importance of the expected flux density variation, shown in Fig. \ref{DRplot}. There is a noticeable phase change $\sim \pi$ and the amplitude ratio $D_{R,j}/D_{R,\bullet} = 0.2 - 3.3$, indicating that for short observation intervals of $\sim$ days--months, flux density variations due to helical motion will dominate over those due to binary orbital motion. 
There will be an additional contribution to $D_{\bullet}$ due to the gravitational lensing of emitted radiation by the orbiting SMBBH for favourable orientations of the SMBBH orbital axis (for $\tilde{i} \sim 90^{\circ}$). For a jet photon emitted close to the SMBBH system with an offset impact parameter $b$, the lensing angle $\tilde{\theta} \sim b/D_L$. The impact parameter is expressed in terms of the photon emission angle $\tilde{\psi}$, using the light bending approximation prescribed by \cite{2002ApJ...566L..85B} \cite[see eqns. (42), (43) and (44) of][]{2015ApJ...805...91M} as
\begin{equation}
b = r \sin \tilde{\psi} \left(1-2 r/r_{\rm g}+(4 r/r_{\rm g}) (1+\cos \tilde{\psi})^{-1}\right)^{1/2}.
\end{equation}
For $r \sim \varpi_0 \sim 147 r_{\rm g}$ and $\tilde{\psi} \in \{0^{\circ},90^{\circ}\}$, ${\rm Max}(\tilde{\theta}) = 0.40 \ \mu$as which is 3 orders of magnitude smaller compared to typical VLBI resolutions of mas indicating that it cannot be resolved. Though, for nearer and less massive objects with beamed emission, this effect is expected to be pronounced.

\section{Conclusion}
\label{conclusion}

The quasar PG 1302$-$102 has been under recent scrutiny owing to the possibility of it hosting a SMBBH system. The focus of the current work is on elucidating its pc-scale jet properties including component kinematics and the variable flux density of the core and components. We obtain observational constraints from the $2 - 8$ GHz and  15 GHz radio data including kinematic parameters and spectral indices. Then, a general relativistic helical jet model \cite[][]{2015ApJ...805...91M} is presented and applied to infer the relevance of the model including a comparison with expectations from a SMBBH scenario. The main results from the study are summarized below:
\begin{enumerate}
\item A mean proper motion $\mu = 0.27 \pm 0.04$ mas yr$^{-1}$, bulk Lorentz factor $\Gamma \geqslant 5.1 \pm 0.8$, inclination angle $i \leqslant 11\fdg4 \pm 1\fdg7$, projected position angle $\lambda = 31\fdg8 \pm 0.2$ and jet half opening angle $\theta_0 \leqslant 0\fdg9 \pm 0\fdg1$ are inferred.
\item A mean spectral index $\alpha = 0.31$ ($2 - 8$ GHz) is inferred.
\item Simulated light curves indicate QPO timescales $\sim$ 10 days or less during the beaming phase with additional observational signatures including their weaker harmonics and power law shaped power spectra.
\item It is inferred that helical signatures could contribute up to $\sim$ 53 percent of the observed core flux density variability. 
\item A case is thus made for high resolution, moderately sampled, long duration ($\sim$ years) observations to reveal helical motion of relativistic knots and to distinguish the expected variable flux density from that expected due to SMBBH orbital activity, which for close separation binaries is expected to appear as a systematic deviation from the helical signature due to accumulative perturbation effects.
\item The distinguishing features include:
\begin{enumerate}
\item A phase difference $\sim \pi$ between $D_{R,j}$ (variable flux density in helical knot scenario) and $D_{R,\bullet}$ (variable flux density in SMBBH scenario).
\item The ratio $D_{R,j}/D_{R,\bullet} = 0.2 - 3.3$ with $D_{R,j}$ dominating the short duration variability ($\sim$ days), especially if observed during the beaming phase.
\item The persistence of the trend $D_{R,\bullet}$ over long durations ($\sim$ years).
\end{enumerate}
\end{enumerate}
The prescription presented can be used to infer helical kinematic signatures from quasars, thus serving as a preliminary kinematics based discriminator, and providing possible candidates for further studies with polarization measurements. The important inference of systematic deviations from this signature arise especially in the context of providing promising candidates for the study of gravitational waves emitted during the inspiral and coalescence of SMBBH systems. With instrumental facilities tuned to the discovery of these signatures in the $(10^{-9} - 1)$ Hz frequency range (PTAs and e-LISA) and with the development and availability of next generation observational facilities such as the upcoming Square Kilometer Array (SKA), there can be a simultaneous monitoring of the electromagnetic counterpart. 


\section*{Acknowledgements}
We thank the anonymous referee for a detailed discussion and offering suggestions which helped improving the manuscript. This work was supported by the SKA pre-construction funding from the China Ministry of Science and Technology. Funding was received from the Hungarian National Research, Development and Innovation Office (OTKA NN110333), and the China--Hungary Collaboration and Exchange Programme by the International Cooperation Bureau of the Chinese Academy of Sciences (CAS). PM thanks CAS for the post-doctoral fellowship supported through the CAS President's International Fellowship Initiative and the National Natural Science Foundation of China for the support through the research grant no. 11650110438. TA thanks the grant support by the Youth Innovation Promotion Association of CAS. PM and TA thank Prof. S. Komossa for discussion and useful inputs to this work. This research has made use of data from the MOJAVE database that is maintained by the MOJAVE team \cite[][]{2009AJ....137.3718L}. The National Radio Astronomy Observatory is a facility of the National Science Foundation operated under cooperative agreement by Associated Universities, Inc. We acknowledge the use of calibrated visibility data from the Astrogeo Center Database of Brightness Distributions, Correlated Flux Densities, and Images of Compact Radio Sources Produced with VLBI. 

\bibliography{bibliography}

\end{document}